# Germanium-Based Mid-Infrared Integrated Photonics

**Delphine Marris-Morini[1], Goran Z. Mashanovich[2], Milos Nedeljkovic[2], Carlos Alonso-Ramos[1], Laurent Vivien[1], Jacopo Frigerio[3], Giovanni Isella[3]**

1. *Centre de Nanosciences et de Nanotechnologies, Université Paris-Saclay, CNRS, 91120 Palaiseau, France*
2. *Optoelectronics Research Centre, University of Southampton, Southampton, SO17 1BJ, UK*
3. *L-NESS, Dipartimento di Fisica, Politecnico di Milano, Polo di Como, Via Anzani 42, 22100 Como, Italy*

*e-mail:delphine.morini@universite-paris-saclay.fr*

**Abstract :**

The mid-infrared (mid-IR) spectral range is a part of the electromagnetic spectrum in which most of the molecules have vibrational and rotational resonances. Ge-based photonic integrated circuits in this wavelength range have thus seen a burst of interest in the recent years, mainly driven by applications related with the detection of chemical and biological substances. Here we review the motivations and the recent developments in this field, from the different material platforms to active and nonlinear devices. We also discuss a few demonstrations of sensing that have already been conducted, attesting the potential applications of such devices. Finally, we conclude by discussing the challenges that have to be solved to transition from lab demonstrations to practical industrial devices.

## 1- Introduction

The images of extremely distant galaxies taken by the James Webb Space Telescope (JWST) give striking evidence of the impact of mid-infrared light on scientific knowledge [1]. While previous telescopes analyzed visible and ultraviolet light, JWST focuses on the infrared part of the spectrum from 0.6 to 28 µm in wavelength. A broader spectral observation of the universe thus provides access to unknown areas, allowing a revolution in celestial observations [2,3]. Back on Earth, in a general context of climate change, hazardous and plastic waste increment and water pollution increase, mid-infrared based photonics can facilitate the transition towards a healthy planet and a new digital world.

The mid-infrared (mid-IR) region is a unique spectral range, defined in a general way from 2 to 20 µm wavelength, hosting the so-called "fingerprint" region (wavelength from 6.7 to 16.7 µm [4]) where most molecules have vibrational and rotational resonances resulting in light absorption at specific wavelengths. Optical absorption spectroscopy, *i. e.* the capability to analyze materials by studying their interaction with a light beam, performed in the mid-IR spectral range is thus an unambiguous way to detect, identify and possibly quantify chemical and biological substances and to perform non-intrusive diagnostics, with a plethora of applications such as detection of small traces of greenhouse or toxic gases, monitoring of industrial emissions or development of new systems for medical diagnostics [5]. In most of these applications, the resolution of the spectroscopic system, which means the minimal width of a spectral feature that can be identified, is a critical figure-of-merit.

Today mid-IR optical systems rely either on bulky table top equipment such as commercially available Fourier Transform Infrared (FTIR) spectrometers, largely used in chemical, physical and biological laboratories [6,7] for precise spectroscopic analysis, or on modules based on the assembly of tunable laser sources such as Interband Cascade Laser (ICL) or Quantum Cascade Laser (QCL) with other optical devices [8,9]. Simple mid-IR optical sensors that combine a broadband light source (e.g. hotplate or LED) with a photodetector (e.g. [10]) are also widely used for gas detection of single species and are cheap and manufactured at volume, but are not sufficiently sensitive for trace gas detection and lack flexibility. In this context, a revolution is still possible, downsizing the dimensions of mid-IR systems and benefiting from on-chip photonic integration. Photonic integrated circuits (PICs) make use of waveguides for routing light between a set of optoelectronic devices, integrated on the same chip, designed to generate, manipulate- and detect light. Developing an efficient photonics platform in the mid-IR range and coupling it with ICL or QCL sources would not only improve the compactness of the systems and their power consumption, but would also bring into play new capabilities triggered by photonic integration such as polarization control, phase or intensity modulation, confinement in a micro-resonator or wavelength conversion through non-linear optical effects.

Different material platforms are used to develop mid-IR PICs, such as III-V semiconductors, chalcogenide glasses or silicon (Si). Si-based photonics already has a major impact in telecom applications operating in the 1.3-1.5 µm wavelength range, and it has been exploited by the main industrial players such as Intel or Cisco [11]. In terms of optical properties, Si photonics benefits from a high refractive contrast between Si waveguide core and silicon dioxide ($SiO_2$) cladding layer which allows tight light confinement in a few hundreds of nanometers-thick Silicon-on-Insulator (SOI) waveguide, providing a compact, high-density and scalable photonic platform. Si-photonics benefits from cutting-edge fabrication tools and large wafer sizes commonly used in microelectronics foundries to achieve high reproducibility (down to nanometers scale) on the fabrication of integrated waveguides [12]. In this context Si photonics is also expected to have a major impact on the development of mid-IR photonics by leveraging the reliable and high-volume fabrication technologies already developed for microelectronic integrated circuits [13]. However, the operation of well-known and mature SOI photonics in the mid-IR is hindered by the strong absorption taking place in the $SiO_2$ cladding beyond 4 µm wavelength [14-16].

To overcome this limitation and benefit from the wide transparency of Si up to 8 µm wavelength, different solutions have been demonstrated to remove $SiO_2$ underneath the waveguide core, exploiting for example subwavelength nano structuration of the waveguide [17]. However, a large part of the mid-IR spectrum is still not accessible, due to the intrinsic loss of Si beyond 8 µm. Interestingly, germanium (Ge), a semiconductor already used in Si-photonics and compatible with Si-technology [18] is transparent in a wider spectral range, from 1.9 to 15 µm [14]. It is thus a promising platform to develop large scale access to mid-IR photonic integrated circuits covering a large part of the mid-IR spectrum, especially the fingerprint region. Since Ge has a large refractive index (around 4 at $\lambda$=8 µm to be compared with that of Si around 3.5 at the same wavelength), high confinement of the electromagnetic field can be expected. Furthermore, the Kerr non-linear index is higher than in most of the materials currently used in non-linear optics, which suggests that efficient non-linear devices can be achieved with Ge-based photonics [19]. Recently, the development of Ge-based mid-IR photonic platforms has witnessed a burst of research activity. Besides the major advantages mentioned above, there are a several challenges related to (i) the choice of a proper cladding material to avoid propagation losses through the evanescent component of the optical mode, (ii) the lattice mismatch of 4.2% between Ge and Si, which is the most-used substrate, and could possibly be responsible for poor material quality and larger propagation losses. Many solutions have thus been proposed in the literature to solve these challenges and to develop Ge-based photonics circuits.

This Review focuses on the role played by SiGe technology in the development of a waveguide-integrated platform comprising both passive (waveguides, couplers, interferometers, resonators) and active devices (photodetectors, modulators heterogeneous integration of laser sources). Emphasis will be placed on non-linear effects (supercontinuum and frequency-comb generation) due to their relevance in sensing applications. It can be noted that the status of the SiGe platform in a broader range of applications and wavelengths has been recently reviewed in Ref. 18, while the state-of-art of GeSn based devices can be found in Ref. 20.

The paper is organized as follows. We start by describing the material and waveguide platforms enabling mid-IR passive devices such as couplers, resonators and interferometers. We discuss the non-linear effects leading to wideband sources through supercontinuum generation in Ge-based photonic platforms. We then report on the development of active devices such as laser sources, modulators and detectors. The progress towards Ge-based frequency-comb generation is then outlined. Finally, we overview the most promising fields of application and summarize of the current state of research in Ge-based mid-IR integrated photonics, together with future challenges and opportunities.

## 2- Material platforms and passive building blocks

**Material platforms.**

Si and Ge form non-polar bonds which inhibit first order interactions between IR photons and phonons [21,22]. As a consequence, the upper limit of their mid-IR transparency window is set by multiphoton absorption taking place beyond ~ 8 µm (*Figure 1*a) and ~ 15 µm (*Figure 1*b) for Si and Ge, respectively [23, 24], where transparency is defined by bulk material propagation loss below 2dB/cm. In epitaxial layers the residual carrier concentration exceeds that of high-resistivity bulk crystals. In particular, in epitaxial Ge a p-type background doping of the order of $10^{15}$ $cm^{-1}$ is typically observed [25], making free carrier absorption a potentially relevant absorption mechanism at longer wavelengths [26].

Interestingly Si and Ge can be assembled in silicon-germanium (SiGe) alloys, whose dielectric constant systematically increases with the atomic number thus enabling light confinement in a Ge-rich core surrounded by a Si-rich cladding layer even with a few percent compositional variation [27]. To take full advantage of the wider mid-IR transparency window of Ge when compared to Si, waveguides with a relatively high Ge content are preferable. Different material stacks and waveguide geometries have been investigated leveraging the high level of maturity reached by SiGe heteroepitaxy and are schematically shown in *Figure 1*.

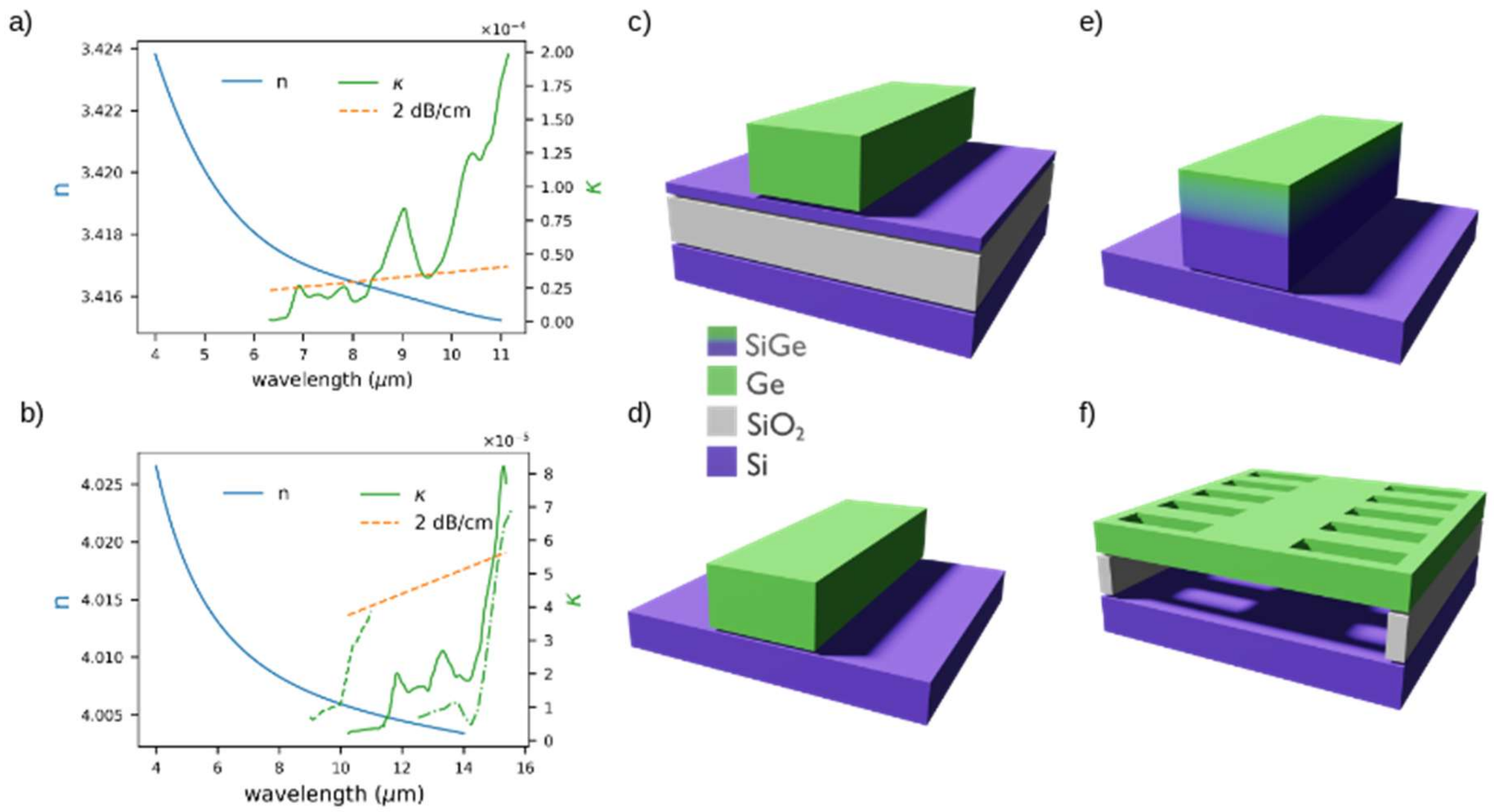

*Figure 1 : Waveguides based on the SiGe material platform. (a) Refractive index n and extinction coefficient κ for Si as reported in Refs. 21 and 23. The dashed orange line identifies the values of κ corresponding to propagation losses equal to 2dB/cm. When k is higher than this limits, losses due to bulk absorption in Si are beyond 2 dB/cm. This line allows to define the transparency limit of Si around*

*8 μm wavelength. (b) Optical constants of bulk germanium. The refractive index n is taken from Ref. 23 the extinction coefficient κ from Ref. 22 (continuous line), 24 (dashed lines) and 28 (dash-dot line). The discrepancy between data reported in the literature can be traced back to the inherent difficulty of measuring relatively weak absorption phenomena. The dashed orange line allows to define the transparency limit of Ge at 15 μm corresponding to data from Ref 24, which is the usually admitted value. Layer stack and schematic representation of different waveguide platforms (c) Ge on insulator d) Ge on Si (e) SiGe graded layers f) suspended Ge.*

The different platforms with their respective advantages and main performances are described hereafter, starting from the popular Ge-on-Si followed by SiGe on Si. Ge on insulator and suspended Ge will then be discussed.

**Ge-on-Si (GOS)**. In a first approach, Ge can be directly deposited on silicon by epitaxial growth. Such Ge-on-Si epilayers, which have been developed for short-wave infrared detection [29], can be grown to the ≈ 2-3 μm thickness required for light confinement. Single mode operation in the mid-IR is then achievable by tuning the waveguide width and etch depth [30, 31]. Several research groups have reported waveguide losses in the GOS platform, the results of which have been plotted versus wavelength in Figure 2. The waveguides had different dimensions and used wafers grown by different sources, with varying recipes for Ge epitaxial growth and waveguide etching, nevertheless we can identify several different features that affect the waveguide propagation losses. Below 2 μm losses are high due to the indirect band edge of Ge, visible in Ref. 32. At wavelengths in the 2-6 μm range the losses are generally low (i.e. <3 dB/cm), with losses <1.2 dB/cm having been reported in [33-35]. At >7.5 μm losses are generally higher than at shorter wavelengths. Ref. 36 showed losses increasing from around 6 dB/cm at 7 μm to >20 dB/cm at 11 μm. Ref. [37] showed lower propagation losses of < 5dB/cm at 7.5-11 μm, which they attributed to the use of GOS wafers that were grown in conditions likely to lead to an n-type background doping (i.e. using an n-type Si substrate and using a growth reactor that had been predominantly exposed to n-type dopants in the past), because free carrier absorption in this spectral range is much lower in n-type Ge than in p–type Ge. In both [36] and [37] the effect of absorption by the Si substrate is also visible: a small peak in loss is visible at 9 μm (corresponding to the Si-O-Si molecular absorption peak that occurs in CZ Si), and in [36] another peak is visible just above 11 μm, corresponding to a multiphonon absorption peak in Si. To minimize this effect thicker Ge top layers could be used to decrease the overlap of the optical mode with the substrate, however thick Ge layers can cause larger thermal strain and wafer bowing. Interestingly, GOS passive waveguide fabrication is now available through a multi-project wafer (MPW) run on 8" wafers (with scanner lithography) from CORNERSTONE [38].

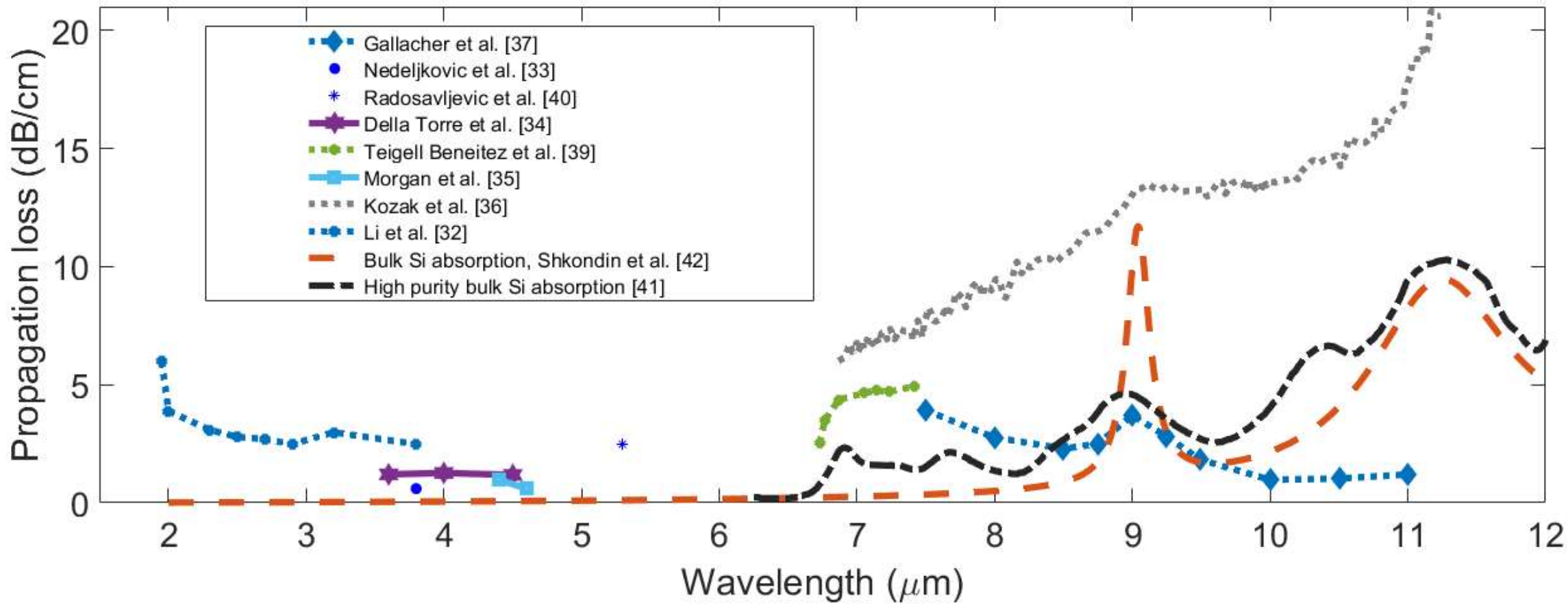

*Figure 2. Summary of reported GOS waveguide propagation loss measurements from literature in the 2-12 µm wavelength range. Curves are also shown for FTIR-measured bulk Si absorption from [41] (where high purity Si wafers were used), and from [42] (where the strong absorption peak at 9 µm that is characteristic of Si-O-Si absorption in CZ Si wafers is visible).*

**SiGe on Si**. SiGe alloys are also largely exploited for mid-IR integrated photonics. Indeed, the refractive index can be engineered by exploiting SiGe alloys instead of pure Ge-on-Si waveguides. In a first approach, SiGe waveguides with a ≈40% Ge content, have been epitaxially grown on Si, with propagation losses of 0.4 dB/cm demonstrated in the 3-4 µm range [43]. In another approach, graded layers, obtained by linearly increasing the Ge content during epi-growth, have been used. In these structures the linear increase of Ge concentration is accompanied by a linear increase of the refractive index, which can be varied to optimize mode confinement (*Figure 1*e). Exploiting rib waveguides featuring a triangular compositional profile, with a maximum Ge content of 40% [44], low propagation loss operation (1-2 dB/cm) over the 3-8 µm spectral range has been demonstrated. Graded SiGe layers have also been extensively used as buffer layers to obtain Ge-rich SiGe alloys on Si substrate (with Ge concentration in the top SiGe layer typically between 80 % and 100%), minimizing the density of dislocations threading to the surface [45]. Such Ge-rich SiGe graded layers present advantages for light propagation beyond 8 µm, as by confining the optical mode in the top part of the graded layer, *i.e*. several micrometers away from the Si substrate, the mode do not suffer from the influence of the multiphoton absorption from the substrate. Propagation losses below 1.5 dB/cm have thus been obtained up to 8.5 µm wavelength, together with losses below 3dB/cm in the 9.5-11 µm range [46]. Besides multiphonon absorption other physical mechanisms might become relevant sources of loss in epitaxially grown SiGe such as residual doping which is hard to control below $10^{15}$ $cm^{-3}$, leading to free-carrier absorption [26]. It is worth noting that Ge-rich graded SiGe waveguides have recently been grown on an industrial-scale 200 mm wafer, with propagation losses below 0.5 dB/cm for 5-7 µm wavelengths and below 5 dB/cm up to 11 µm [47] which establishes a foundation for a scalable, silicon compatible mid-infrared platform accessible in industrial foundries.

**Ge-on-insulator (GOI)**. Similar to the most popular Si-based photonics platform, that of Si-on-insulator (SOI), GOI can be realized by the smart-cut technique [48,49], or by thinning down the SOI to a very thin Si top layer and subsequently growing Ge on top of this thin SOI substrate [50] (*Figure 1*c). The main motivation for the utilization of this platform can be the realization of high-speed active devices because the carrier mobility in Ge is larger than in Si. In terms of mid-IR light guiding, propagation losses of 2.3 ± 0.2 dB/cm were demonstrated on an n-type GeOI wafer at a wavelength of ~2 µm [49].

However, targeting longer wavelengths, GOI suffers from the same limitation as SOI, namely the buried oxide layer limits the operation in the mid-IR up to ~ 4 μm. Nevertheless, as a main advantage, GOI offers the possibility to suspend Ge waveguides by etching the buried oxide, to utilise the entire transparency range of Ge. Suspended Ge platforms will be discussed below. Furthermore, when oxide is replaced with mid-IR transparent materials ($Al_2O_3$, $CaF_2$, ZnSe etc) the operational wavelength range of the platform can be significantly enlarged [51,52].

**Suspended Ge**. In the mid-IR, Ge is transparent up to 15 $\mu$m wavelength, however, Si is transparent only up to 8 $\mu$m wavelength. Hence, the operation of Ge-on-Si technology is strongly limited by the absorption of silicon. This limitation has been overcome by implementing suspended Ge waveguides [53] (*Figure 1*f). The starting material can be either GOS or GOI. Holes around the waveguide core can be dry etched to give access to a liquid or vapor phase that can etch the oxide below and subsequently suspend the waveguide core. For GOS such an etchant is TMHA (Tetramethylammonium hydroxide) or $XeF_2$, whilst for GOS it is HF (hydrofluoric acid), as for suspending Si [54]. In terms of the holes, they can be etched at a relatively large distance from a rib waveguide to not disturb the optical mode, or they can be etched next to the core but designed such that they do not cause any reflection or scattering and the mode will see them as an average refractive index. The latter approach is using the subwavelength gratings (SWG) as a lateral optical cladding. Indeed, periodic nanostructures with subwavelength period provide unique degrees of freedom to control the propagation of light in integrated devices [55,56]. By using a period shorter than half of the wavelength, diffraction and reflection effects are suppressed. Hence, the periodic structures behave as a homogeneous medium with properties like modal confinement or dispersion being controlled by the geometry of the periodic lattice. This approach has been extensively developed in recent years, for example, for the implementation of all-dielectric metasurfaces [57,58]. Subwavelength nanostructures have been successfully exploited to develop high-performance waveguide-based devices [59-61], including foundational demonstrations of fiber-chip couplers [62], antireflective waveguide facets [63] and lenses [64]. Pioneer demonstrations show that subwavelength metastructures can also play a key role in the development of high-performance Ge-on-Si and suspended Ge photonic devices for applications in the mid-IR.

Suspended Ge waveguides with lateral subwavelength grating comprise a solid Ge core surrounded by two subwavelength lattices that play a triple role, as they provide a lower effective index to confine the optical mode at the waveguide core, allow the penetration of hydrofluoric acid to remove the oxide underneath, and provide mechanical stability once the underneath layer has been removed. Furthermore, these structures present two advantages over the former approach (where holes are etched far from the waveguide to avoid mode perturbation): 1) two dry-etch steps need to be carried out to initially define a Ge rib waveguide, and then to create an array of holes alongside the waveguide whilst in the SWG case only one dry-etch step is required; 2) the SWG waveguide uses the entire thickness of the top Ge layer which gives better mechanical robustness. Both waveguide types have been characterized at a wavelength of 7.67 um with losses of 2.5 dB/cm for the rib [30] and 5.3 dB/cm for the SWG structure [53]. Future work should focus on improving the loss and characterization at longer wavelengths.

## Passive building blocks

A number of passive building blocks have been developed in all these Ge-based material platforms and waveguides. A few selected results will be discussed in this section.

Mid-IR light coupling to Ge-based waveguides can be achieved both by in plane and out of plane couplers. For in plane coupling (butt coupling) usual approaches involve polishing or cleaving, which do not always give high quality Ge facets. An alternative approach is ductile dicing. Dicing is often used to mechanically separate chips from wafers. Dicing can be used to produce optical grade surfaces with vertical sidewalls and nanoscale roughness by performing the dicing in the ductile regime, whereby the strain applied during machining is kept within the material's plastic limit [30]. This technique employs stringent machining parameters for blade grit size and grit concentration, blade rotational speed, sample translation speed, depth of cut, and sample coolant; all must be optimized and maintained to sustain ductile machining. When the strain applied during machining exceeds the material's plastic limit, brittle machining is instead initiated, which results in a rougher, cracked, and chipped cut sidewall. Grating couplers are standard devices in Si-photonics and there is a range of different designs showing good coupling efficiencies. Ge-on-Si waveguides entail significant challenges for the implementation of fiber-chip grating couplers due to a moderate index contrast between the Ge waveguide and the Si substrate ($\Delta n \sim 0.6$) which substantially limits the directionality and strength of the couplers. Fiber-chip grating couplers for the Ge-on-Si technology were first demonstrated in 2015, working at a wavelength of 3.8 $\mu$m [33]. Ge-on-Si waveguides have a comparatively large refractive index that results in strong back-reflections at the waveguide-to-grating interface. Low reflectivity is particularly important for spectroscopic applications where the absorption features with a bandwidth of a few hundreds of megahertz need to be precisely resolved. To solve this issue a single-etch Ge-on-Si grating coupler with an inversely tapered access stage, operating at a 3.8 µm wavelength has been demonstrated. The coupling efficiency was 7.9% and the reflectivity was below −15 dB (3.2%) [65]. They also showed that a focusing geometry allowed a 10-fold reduction in inverse taper length, from 500 to 50 µm and improvement in directionality of the grating from 30% to 60%. The coupling efficiency could be increased by optimizing the etching depth, apodizing the grating, or using a bottom mirror. A grating coupler has been designed for suspended Ge based on a micro-antenna design with only three radiative elements and an adaptation section to reduce the reflection. A coupling efficiency of ~ 40% and a 1-dB bandwidth broader than 430 nm around a wavelength of 7.7um were reported, which is almost twice the typical fractional bandwidth of a conventional grating coupler [66]. In addition, the proposed design was tolerant to fiber tilt misalignments of ±10°. The main issue was a thin adaptation section which, due to strain, could break during the undercutting of the Ge waveguide.

Ring and racetrack resonators are crucial devices for filtering, non-linear applications, optical modulation or sensing. One of the most important resonator parameters is the Q-factor. High Q-factor SiGe-based resonator has been recently demonstrated in the 3.5–4.6 µm with quality factors reaching up to one million [67]. Shifting to longer wavelengths, it is challenging to achieve similar levels of performances mainly because the Q-factor scales with the inverse of the wavelength. Nevertheless, suspended Ge disk resonators have demonstrated loaded Q-factors of 224,000 at 8 µm wavelength [68], however coupling waveguides still have to be integrated. SiGe graded-index waveguides were also used to demonstrate racetrack resonators operating in the long-wave infrared spectral range from 7.5 to 9.0 µm, with a quality factor up to 100,000 [69].

1D nanocavity-based resonators have been reported in suspended Ge at $\lambda$ = 2.5 µm with a Q-factor of 18,000 [70], as well as a Bragg grating Fabry-Perot cavity in the 6µm thick SiGe graded platform with a maximum Q-factor of 2200 at a wavelength of 7.95 µm [71]. The Bragg grating structure was based on a top-surface waveguide corrugation that provided a rejection higher than 20 dB.

Arrayed Waveguide Gratings (AWGs) and Planar Concave Gratings (PCGs, also called echelle gratings) are multiplexers / demultiplexers that can be used for mid-IR communications or as the core elements in mid-IR spectrometers. Both have been demonstrated in a 2 µm GOS platform at a wavelength of ~5 µm. A 200 GHz channel spacing 5-channel AWG with an insertion loss/crosstalk of -2.5/-3.1 dB and -20/-16 dB for TE and TM polarization, respectively, was demonstrated. GOS waveguide losses in the range 2.5–3.5 dB/cm for TE polarized light and 3–4 dB/cm for TM polarized light in the 5.15–5.4 µm wavelength range were reported [72]. 6-channels PCGs based on the same GOS platform and operating in the same wavelength range, with two different types of gratings (flat facet and distributed Bragg reflectors) were analyzed for both TE and TM polarizations. The insertion loss and cross talk for flat facet PCGs were found to be -7.6/-6.4 dB and -27/-21 dB for TE/TM polarization. For distributed Bragg reflector PCGs the insertion loss and cross talk are found to be -4.9/-4.2 dB and -22/-23 dB for TE/TM polarization [73].

Miniaturized mid-IR spectrometers providing broadband operation and fine resolution have an immense potential for applications in sensing. Fourier-transform spectrometers (FTS) are of particular interest for on-chip integration. To achieve both large bandwidth and small resolution spatial heterodyning has been combined with optical path tuning by the thermo-optic effect. The high resolution was provided by spatial multiplexing among different Mach-Zehnder interferometers with increasing imbalance length, while the broadband operation was enabled by fine tuning of the optical path delay in each interferometer harnessing the thermo-optic effect. Taking advantage of the graded SiGe platform a mid-IR SiGe FTS has been experimentally demonstrated, with a resolution of 13.4 $cm^{-1}$ and a bandwidth of 603 $cm^{-1}$ near 7.7 µm wavelength with a 10 MZI array [74]. This resolution approaches that of widely used commercial benchtop FTIR spectrometers—for example the Agilent 5500 compact FTIR spectrometer [75] has a specified resolution of 2-16 $cm^{-1}$.

## 3- Non-linear effects in Ge-based photonic circuits for wideband optical sources on chip

Nonlinear optics covers a wide range of phenomena related to the non-linear response of the polarization density to the electric field of the light. These phenomena are usually triggered by short pulse-laser sources, providing high peak intensity. Motivations to develop nonlinear optics cover both fundamental physics studies and more applied innovations, requiring the spectral conversion of light to different regions of the electromagnetic spectrum. In the context of mid-IR absorption spectroscopy, the possibility to obtain wideband light sources is of particular importance to permit the interaction of light with complex molecules or with closely spaced absorption peaks of a given molecule. Being centrosymmetric materials, non-linear effects in Si and Ge are based on the third order susceptibility $\chi^{(3)}$ or on the related non-linear refractive index ($n_2$). Different photonic integrated platforms have been compared in terms of nonlinear properties in [76]. Pure Ge possesses among the highest $n_2$ values (expected value up to 2.5×$10^{-13}$ $cm^2$/W in the mir-IR [77]), higher than Si (6×$10^{-14}$ $cm^2$/W), and much higher than $SiO_2$ (2.7×$10^{-16}$ $cm^2$/W) and silicon nitride (2.4×$10^{-15}$ $cm^2$/W). It competes mainly with AlGaAs and InGaP (~$10^{-13}$ $cm^2$/W) and chalcogenides (9×$10^{-12}$-9×$10^{-14}$ $cm^2$/W). When looking at the non-linear refractive index of SiGe alloys, it can be seen that its non-linear refractive index is not expected to be linearly proportional to Ge and Si concentration in the alloy. For Ge concentration below 80%, it is predicted that the maximum value of $n_2$ saturates at $10^{-13}$ $cm^2$/W, while it increases to 2.$10^{-13}$ $cm^2$/W for $Si_{0.1}Ge_{0.9}$, close to the value of pure Ge [77]. This evolution is explained by the different bandstructure of $Si_{1-x}Ge_x$ alloys between a Si-like behaviour (0 ≤ x ≤ 0.8) and a Ge-like behaviour (0.8 < x ≤ 1) [77,78]. To achieve efficient non-linear conversion, not only is a large non-linear refractive index required, but also a good mode confinement within the non-linear medium is essential. This is evaluated through the nonlinear parameter $\gamma$ defined as $\gamma=\frac{\omega_0\, n_2}{c\, A_{eff}}$ where $A_{eff}$ is the effective area of the waveguide mode$A_{eff}\omega_0$. Besides the nonlinear properties of the waveguide mode, non-linear processes also require low propagation losses, and modal dispersion engineering which is a key parameter to control short pulse propagation within the waveguide. Different approaches have been undertaken to optimize these parameters. Most of the reports on non-linear processes in Ge-based mid-IR photonics focus on the demonstration of Supercontinuum Generation (SCG) which is defined as an extreme spectral broadening of an input pulsed laser light [76]. SCG has been reported in different SiGe waveguides configurations [79-83, 43] as well as in Ge-on-Si waveguides [34]. An extensive review of these results can be found in [76]. Among the main results, one octave SCG was generated on chip from 3 to 8.5 µm wavelength in an air-clad step index $Si_{0.6}Ge_{0.4}$ waveguide [43] (*Figure 3*,(a)-(b)) while on-chip two octave generation was obtained from 3 to 13 µm wavelength using Ge-rich graded index SiGe waveguides [83] (*Figure 3*,(c)-(d)). Furthermore, dispersion engineering along the propagation direction has been proposed to enhance SCG, especially the spectral flatness and coverage [84] (*Figure 3*,(e)-(f)). While being impressive in terms of spectral conversion achieved within the photonics circuits, all these demonstrations have been performed using table-top high-power, wavelength-tunable femtosecond laser. Achieving a compact integrated photonics platform requires the supercontinuum generation to be triggered by compact laser sources. The generation of ultrashort and intense pulses in the mid-IR range from cascade lasers is currently a subject of high interest in the community [85]. As a few examples, around 8 µm wavelength, pulses have been generated using active mode locking [86], while external pulse compression of frequency modulated frequency comb-QCL has shown the possibility to achieve femtosecond pulses with peak powers up to 4.5 W [87]. Interestingly, Watt-level optical pulses have also been obtained by direct high frequency modulation of the pump current [88]. Recently, chirped Bragg gratings based photonic circuit, engineered to compensate for the group delay dispersion of quantum cascade laser frequency comb sources have been exploited to generate on chip pulses as short as 1.39 picoseconds [89]. Picosecond and sub-picosecond pulses have also been generated from ICL at 3.6-3.8 µm wavelengths [90,91]. All this recent progress is promising for the integration of a compact pulse pump source to trigger SCG in SiGe photonics circuits. Among the remaining challenges, the efficiency of the SCG process has to be optimized to be compatible with the maximum power delivered by the pulsed laser. A dispersion engineered waveguide made of two sections, one providing strong anomalous dispersion and the second allowing propagation in a low dispersion regime, has been used to obtain SCG at low pump power [92]. Wideband supercontinuum has been obtained with a few hundred Watts peak power at the input (Fig 3, (g)-(h)), which is typically one order of magnitude lower than previous demonstrations. Such progress has to be pushed forward in the near future, as there is still a discrepancy between the pulsed cascade laser power and the pump power required to trigger on-chip SCG.

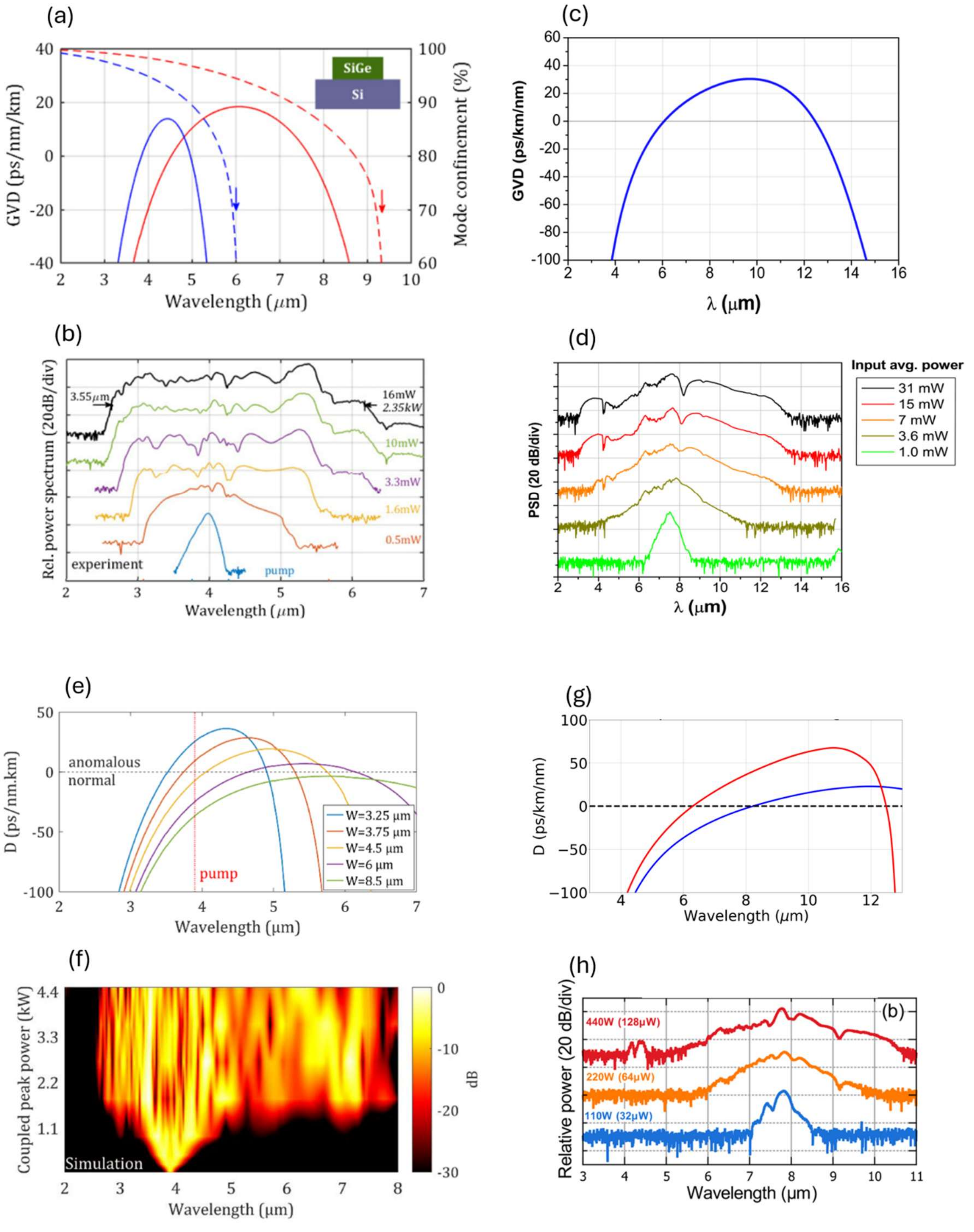

*Figure 3. SCG generation in SiGe waveguides. (a) Dispersion in air-clad step index $Si_{0.6}Ge_{0.4}$ waveguide (b) SCG generation for different input power in air-clad step index $Si_{0.6}Ge_{0.4}$ waveguide; (c) Dispersion in graded SiGe waveguide; (d) SCG in graded SiGe waveguide; (e) Dispersion in a two-step SiGe/Si waveguide for dispersion engineering along the propagation; (f) SCG in the two step SiGe/Si waveguide. Panels (a) and (b) adapted from [43]. Panels (c) and (d) adapted from [83]. Panels (e) and (f) adapted from [84]. Panels (g) and (h) adapted from [92].*

## 4- Active devices

The emergence of Ge-based waveguide platforms has allowed research into optoelectronic integration of active devices for light generation, modulation, and detection to begin in the last few years, as a key step towards realizing fully integrated mid-IR PICs.

GeSn offers some potential for monolithic integration of group-IV lasers and photodetectors, since incorporating Sn into Ge and applying strain to it can create a direct bandgap in the 2-4µm range [93]. Optically pumped lasing of GeSn has been shown at a temperature of 270 K with emission in the 2.8-4 µm range [94], and electrically injected lasing has been shown at 100 K with emission at 2.3 µm [95,96]. However, for the time being the emitted power is low, and even in mature devices GeSn will not reach the crucial fingerprint region for mid-IR spectroscopy. Excitingly, epitaxial growth of III-V QCLs on Si substrates has been achieved with performances comparable to similar devices grown on native substrates [97]. Waveguide integration strategies will need to be developed so that they can be manufactured alongside and coupled to Ge-based PICs. High performance waveguide integrated lasing in the mid-IR can also be achieve through heterogeneous integration of ICLs and QCLs. [98] has recently demonstrated flip-chip bonded integration of Ge-on-Si waveguides and QCLs. Room temperature light emission was observed from Ge-on-Si waveguides that were butt-coupled to Fabry-Perot QCLs. This approach allows separate manufacturing of III-V lasers and Ge-based PICs using ideal processes but requires careful alignment of the laser and waveguide chips at the back end of line.

Interestingly, mid-IR photodetection has been achieved in SiGe/Si stacked heterojunctions through free-carrier absorption [99], as well as in SiGe/Si quantum wells via intersubband absorption [100]. More recently photodetection has also been observed at room temperature in graded index SiGe Schottky junctions between 5 µm and 8 µm [101] and PIN junctions between 5.5 µm and 10 µm [102]. The photocurrent generation mechanism in both devices is not fully understood—it may arise from defect-mediated sub-bandgap absorption, caused by defects resulting from misfit dislocations in the SiGe. These devices may already be sensitive enough for some signal monitoring and sensing applications, and a deeper understanding of the photodetection mechanism may yield further improvements. Bolometers (and other types of thermal photodetectors) could also be integrated with Ge-based waveguides to create room temperature devices with CMOS-compatible fabrication processes and may particularly be of use in applications where the lowest noise equivalent powers and highest bandwidths are not required. Interestingly, it has been predicted [103] that nanophotonic engineered thermal photodetectors have the potential to achieve better noise performance at wavelengths above 5 µm than all other types of photodetectors. A waveguide integrated photodetector using bolometric effect has been demonstrated recently, achieving a noise-equivalent power of $4.03 \times 10^{-7}$ $W/Hz^{0.5}$ at 4180 nm [104].

In parallel, there are several alternative routes towards the integration of photodetectors in Ge and SiGe PICs. Heterogeneous integration of III-V photodetectors (e.g. quantum cascade detectors, type II superlattice detectors, etc) would enable integration of high performance devices, using flip-chip bonding or transfer print techniques. 2D materials (such as graphene, black phosphorus, and $WS_2$/$HfS_2$) have been used to implement photodetectors integrated with Si waveguides [105] at wavelengths from the near-IR up to 7.1 µm. Such devices could be transferred into Ge platforms.

Near-IR SOI photonics have driven exhaustive research and optimization of high speed near-IR modulators. Even though Si does not have a linear electro-optic effect, SOI PN junction modulators based on the free-carrier plasma dispersion effect have been able to reach on-off keying modulation rates of 112 Gbit/s with co-packaged driver electronics and power consumption of 1.59 pJ/bit [106]. In the mid-IR modulation is interesting both for sensing applications and for free-space optical communications in the mid-IR atmospheric transparency windows. For example, modulation may be used for frequency comb generation for dual comb spectroscopy, for synchronous detection for noise-reduction in absorption spectroscopy, for switching as part of referencing schemes, and to implement time delays in spectrometers. Ge is a centrosymmetric crystal, like Si, so it does not have a linear electro-optic effect. As in near-IR Si photonics, the most direct methods for modulating light in Ge are to either exploit the thermo-optic effect for phase modulation at kHz frequencies, or to use the free-carrier plasma dispersion effect for higher frequency phase or intensity modulation [107]. Both effects can be employed throughout the mid-IR.

Although Ge has a strong thermo-optic coefficient, thermo-optic modulation is challenging in the Ge-on-Si and SiGe-on-Si waveguide platforms due to the high thermal conductivity of the Si substrate, which allows heat to easily escape from the waveguides. Ref. 108 has shown that the efficiency can be improved by either locally suspending the waveguide near the heater, or by incorporating a buried oxide layer into the substrate (positioned far enough below the waveguide so as not to interact with its optical guided mode).

Modulation by the free-carrier plasma dispersion effect is both faster and more efficient. Changing the free-carrier concentration in Si and Ge modifies both the real refractive index ($\Delta n$) and absorption coefficient ($\Delta\alpha$) of the materials. Both $\Delta\alpha$ and $\Delta n$ generally become larger with increasing wavelength throughout the mid-IR, increasing approximately as the square of the wavelength [109,26], however, the size of the optical mode generally increases with the wavelength, which can limit the overlap of the optical mode with the region where free carrier concentration varies and reduce the modulator efficiency. Furthermore, Ge has a stronger $\Delta\alpha$ than Si, and a similar $\Delta n$. In both SiGe and Ge, the result is that in the mid-IR any phase modulation is accompanied by a significant excess absorption change, and that intensity modulators are most effective when achieved directly through absorption modulation. Translating near-IR SOI free-carrier modulator designs into high frequency Ge-based mid-IR waveguide modulators is not trivial, because of the wavelength scaling of the free-carrier effect, and because Ge and SiGe waveguides have widths and heights on the order of a few micrometers (compared to hundreds of nanometers in SOI) thus requiring relatively large space-charge regions. Free-carrier absorption and phase modulators have been achieved in GOS [110] as shown in Fig. 4 (a), with laterally positioned waveguide integrated PIN diodes operated in the carrier injection regime at 3.8 µm wavelength, reaching extinction ratios of up to 35 dB, and showing modulation at 60MHz. Similar modulators were shown to work under DC conditions at 8 µm wavelength, with a lower 2.5 dB extinction ratio. Extremely broadband graded index SiGe modulators have reached 1 GHz modulation across the 5-9 µm wavelength range [111] by integrating a vertical Schottky diode with the SiGe waveguide (Fig. 4 (b)) and operating it under reverse bias to vary the size of the resulting depletion region. The device geometry was engineered such that a depletion region created by placing a metal contact directly on top of the waveguide had a good overlap with the optical mode. A vertical PIN diode integrated with SiGe waveguides was finally demonstrated in [102] at 5.5-10 µm with N+ and P+ layers introduced into the waveguides using in situ doping (Figure 4(c)). This modulator was able to reach modulation up to 1.5 GHz, with improved extinction ratio reaching up to 3.2 dB at 10 µm with depletion and 10 dB at 10 µm for injection.

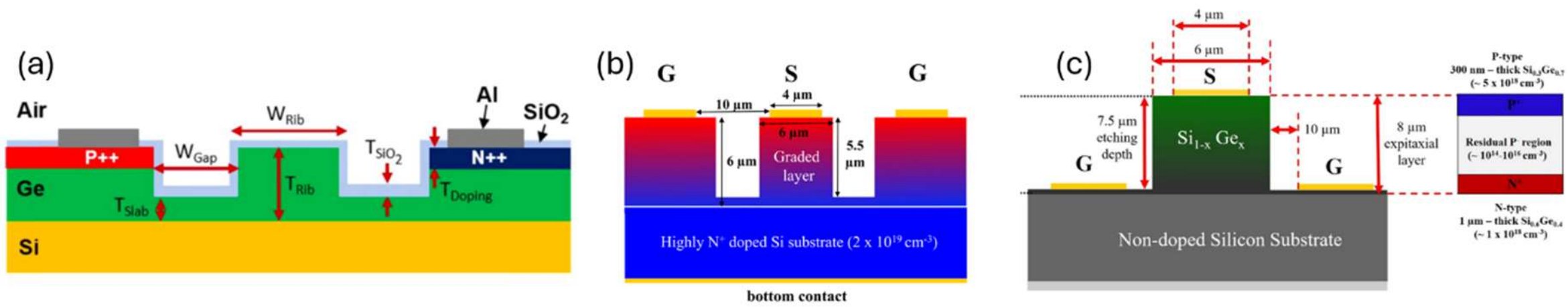

*Figure 4. Mid-IR integrated electro-optical modulators : schematic views of (a) the GOS modulator exploiting a lateral PIN diode reported in [110], the graded SiGe modulators exploiting (b) a Schottky diode as reported in [111]; (c) a PIN diode reported in [102]. All these modulators exploit free carrier concentration variations induced by the applied voltage to modulate the light phase and amplitude.*

Other novel designs for Ge-based platforms that exploit other modulation mechanisms have been proposed through theoretical simulations, based on the quantum cascade Stark effect in Ge/SiGe quantum wells [112] and GeSn/SiGeSn quantum wells [113], or on graphene integration [114,115]. These devices may in future help to extend mid-IR modulation up to 10s of GHz bandwidths.

## 5- <u>Towards frequency comb generation</u>

Optical frequency combs (OFC) have garnered significant interest and have become invaluable tools in various scientific and technological domains [116]. In the mid-IR spectral range, OFCs have a huge impact, thanks to their ability to bridge optical and RF spectral domains. Furthermore, dual-comb spectroscopy is a unique tool that allows the optical spectrum to be measured with high frequency resolution, with large impact in the area of optical sensing. Different techniques can be employed to generate optical frequency combs, and first demonstrations relied on mode locked lasers. Indeed, the phase coherent train of ultra short pulses at the output of the mode locked laser translates in the frequency domain into a set of regularly spaced optical lines, whose spacing is driven by the cavity optical round-trip time. Frequency combs generated from QCL and ICL have thus been demonstrated [85-91].

In parallel, micro-resonator-based comb sources are a subject of huge interest and development. The non-linear Kerr effect can thus be exploited in a resonant cavity to enhance light matter interaction. Optical parametric generation and cascaded four-wave mixing occur in the resonator, and with an appropriate dispersion, an optical frequency comb can emerge at the resonator output, whose line spacing is given by the cavity free spectral range. There has been a huge amount of work dedicated to microresonator-based comb generation in the near infrared range [117,118], and conversion efficiencies exceeding 50% have been recently obtained by [119].

In the mid-IR range, demonstrations have been reported, relying on both Silicon nitride and Silicon on insulator waveguides[120-122]. However all these demonstrations are limited in the short wavelength part of the mid-IR, and there is currently no integrated micro-resonator frequency comb beyond 5 µm wavelength. The main reason for this is the difficulty of demonstrating high finesse resonators at this wavelength range. SiGe-based photonics circuits being a candidate of choice for mid-IR photonics, high performance resonators are currently under investigation. However, resonators showing a Q-factor higher than $10^5$ have been reported recently up to 8µm wavelength [68-69] paving the way for future micro-resonator based comb generation in the long wavelength part of the mid-IR spectrum.

Finally, electro-optical frequency combs provide an attractive alternative approach for generating tunable frequency combs [123]. In this case, the periodic modulation of the phase or amplitude of a light beam transfers into the generation of spectral lines in the frequency domain, separated from the optical carrier by the modulation frequency. Interestingly this method brings tunability in the line spacing, that can be easily tuned by changing the repetition rate of the RF electrical signal applied on the modulator. Following the recent demonstration of an integrated Schottky-based SiGe modulator [111] , a mid-IR electro-optical frequency comb has been recently demonstrated [124]. As shown in Fig. 5, more than 100 lines have been generated around the optical carrier at 8 µm wavelength, spanning over 2.4 GHz. Such a result already demonstrates integrated and tunable mid-infrared electro-optic frequency-comb generation systems on SiGe photonics circuit.

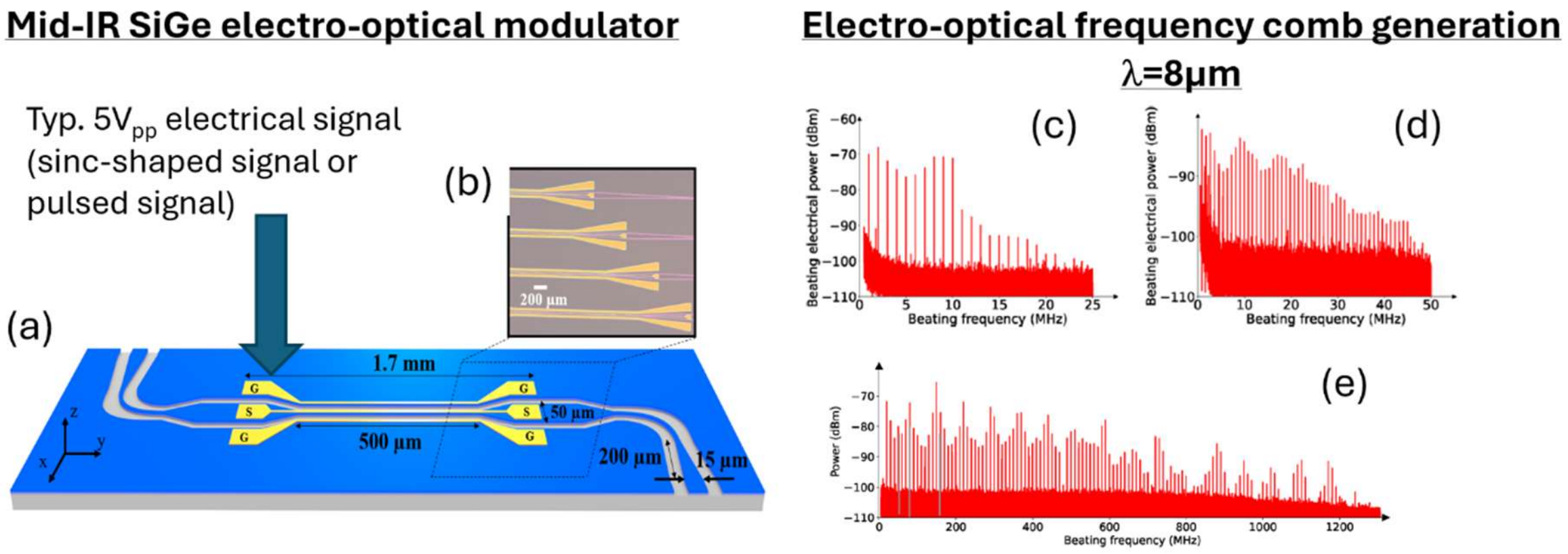

*Figure 5 : Electro-optical frequency comb generation in the mid-IR (a) schematic view of the integrated electro-optical modulator; (b) microscope image of the fabricated device; beat note measurement when (c-d) a sinc-shape periodic electrical signal is applied on the modulator (e) a pulsed electrical signal is applied on the modulator. Adapted from [124]*

## 6- Applications of Ge based mid-IR photonics

The main motivation to develop integrated photonics in the mid-IR is related with sensing. Interestingly, there have been already a few demonstrations of sensors that utilized Ge-based material platforms. In 2013, P. Wagli et al. [125] showed that GOS waveguides can be used to detect cocaine in human saliva. They combined a microfluidic droplet-based liquid−liquid extraction method to transfer cocaine continuously from MIR-light-absorbing saliva to a MIR-transparent solvent (tetrachloroethylene). The lowest detected concentration that authors achieved was 100 µg/mL. They used a QCL operating at 5.71um (1750 $cm^{-1}$) and a MCT detector. They argued that this relatively high limit of detection could be reduced by improving liquid-liquid extraction, temperature stabilization of the laser, and better coupling of the light source to the GOS waveguide. V. Mital et al [126] used a much simpler fluidic delivery on GOS multimode waveguides, filter paper as a flow strip. Absorption spectroscopy of bovine serum albumin (BSA) in phosphate buffered saline (PBS) was experimentally demonstrated in the 5.3−10.0 µm (1900−1000 $cm^{-1}$) region. Multimode GOS waveguides were guiding light up to a wavelength of 12.9 µm. Limit of detection was 100 µg/mL and the amide I, II, and III peaks were observed and quantified. Finaly a novel GOS waveguide design, which basically represented a Ge-based attenuated total internal reflection (ATR) unit on-a-chip, was used for sensing toluene in water [39]. The 2 µm thick GOS waveguide operated in the 6.5-7.5 µm wavelength range. Monolithically integrated microlenses on the back of the chip in combination with Au-coated grating couplers provide a broadband and tolerant optical interfacing. Preconcentration of toluene was achieved by a 3D hexagonal mesoporous silica coating which was modified to be hydrophobic, resulting in a LoD of 7 ppm.

Related with such practical sensing applications, the question of which cladding to use arise rapidly. Indeed, most of the device demonstration rely on air-cladded waveguides. However, claddings are needed to protect the surface of the chip, for opening sensing windows to enable light-analyte interaction in photonic sensors, or for isolation of waveguide modes from metal contacts or microfluidic channels for example. Furthermore, since Ge native oxide is weak and water- soluble it can be interesting to add a protective layer especially when liquid analytes are being detected. In the near-IR, silicon dioxide ($SiO_2$) is usually used as the cladding for SOI photonic circuits due to its excellent transparent, chemical and mechanical properties and ease of deposition. However, beyond 4μm, $SiO_2$ cannot be used due to high material loss and therefore alternative cladding materials need to be considered. The criteria that potential cladding materials must meet are threefold: (1) transparency in the mid-IR; (2) compatibility with scalable fabrication processes; and, (3) deposition with sufficient thickness to optically isolate the waveguide without delamination. Potential candidates are amorphous Si (a-Si), $HfO_2$, $Ta_2O_5$, $Nb_2O_5$, $CaF_2$, ZnSe, $ZrO_2$, $Al_2O_3$, $TiO_2$ or Teflon and other polymers. For passivation of Ge surface in a plasmonic mid-IR sensor, atomic layer deposition (ALD) can be used, as shown in [127]. The authors have used ~10nm thick layers of $ZrO_2$, $Al_2O_3$, $TiO_2$ and $HfO_2$. The first two provided already protection of the plasmonic waveguides for 90 min in direct water exposure. The latter two showed inferior performance but it can be expected that by increasing their thicknesses a similar level of protection could be achieved. For thicker layers other techniques can be used such as evaporation, or sputtering. A recent paper [35] investigated the propagation loss of GOS waveguides at 4.4-4.6 µm wavelength when cladded with several different materials, finding the lowest excess loss when $Nb2O_5$ was used. This area of research has been slightly neglected and further investigation of suitable cladding materials at wavelengths beyond 4.6 µm will need to be adequately addressed if we want to see the realisation of commercial Ge-based mid-IR sensors.

## 7- Outlook and perspective

Silicon Photonics has made a huge progress in the last 20 years and is currently having a major industrial impact in the telecom area, its extension from near-IR to mid-IR wavelength range, enabled by Ge-based platforms is expected to have large repercussion in many areas, starting from sensing applications. Considering the recent evolution of the field, compact and high performances spectrometers can be envisioned in the near future. A solution for fast and accurate measurements is based on the development of dual comb spectrometers [128], where two optical frequency combs with slightly different repetition rate are exploited, allowing to map the optical spectrum in the RF domain. It will be very beneficial if such spectrometers are demonstrated in the coming years.

In this context, several scientific and technological challenges can be identified for the coming years, defining the main research priorities in this field:

- The laser source integration needs to transition from lab demonstrations to practical industrial devices. Coupling Ge-based waveguides with efficient electrically driven sources is one of the challenges to be addressed. Interestingly, monolithic and heterogeneous integration of III-V QCLs or ICLs are both improving rapidly, and results can be expected in the short term.
- Ultra-wide band supercontinuum sources have been demonstrated on chip, by harnessing optical nonlinearities on chip. Impressive results have already been obtained, such as the possibility to generate one or two octaves on chip, but progress is still required to make these devices compatible with compact cascade laser sources as optical pumps.

- Improving the performance of high speed amplitude modulators, especially the efficiency/loss ratio is another major challenge that can drastically push Ge-based mid-IR photonics, either to improve electro-optical frequency comb generation or to pave the way for free space communications applications. Furthermore, up to now, high speed modulators rely on amplitude modulation only. The development of phase modulation would be an important result, however there is no practical path in this direction at the moment.
- Significant further efforts are also required to achieve frequency comb generation based on resonators in the mid-IR range [129].
- Finally, to show the impact of mid-IR photonics in everyday life, sensing demonstrations will have to be pushed forward to demonstrate high sensitivity detection of multiple molecules simultaneously.

**Acknowledgements**

D. Marris-Morini, C. Alonso-Ramos, L. Vivien, J. Frigerio, G. Isella acknowledge support from the ANR (Light-Up Project - ANR-19-CE24-0002-01), from the European Union's Horizon Europe ERC program (ERC-Electrophot, 101097569) and from the European Union's Horizon Europe research and innovation programme (UNISON – 101128598). D. Marris-Morini, C. Alonso-Ramos, L. Vivien acknowledge support from the RENATECH network. G. Z. Mashanovich and M. Nedeljkovic acknowledge support from the EPSRC programme grant MISSION (EP/V047663/1).

**References**

[1] Rieke, G., Wright, G. A mid-infrared dream come true. Nat Astron 6, 891 (2022).

[2] Levan, A.J., Gompertz, B.P., Salafia, O.S. et al. Heavy-element production in a compact object merger observed by JWST. Nature 626, 737–741 (2024).

[3] Bell, T.J., Welbanks, L., Schlawin, E. et al. Methane throughout the atmosphere of the warm exoplanet WASP-80b. Nature 623, 709–712 (2023).

[4] Infrared Spectroscopy: Fundamentals and Applications, Stuart B., 2004 John Wiley & Sons, Ltd ISBNs: 0-470-85427-8 (HB); 0-470-85428-6 (PB).

[5] Beć, K.B., Grabska, J., Huck, C.W., Biomolecular and bioanalytical applications of infrared spectroscopy – A review, Analytica Chimica Acta, 1133, 150-177 (2020).

[6] Valand, R., Tanna, S., Lawson, G., Bengtström, L., A review of Fourier Transform Infrared (FTIR) spectroscopy used in food adulteration and authenticity investigations, Food additives & contaminants: Part A, 37:1, 19-38 (2020).

[7] Bacsik, Z., Mink, J., Keresztury, G., FTIR spectroscopy of the atmosphere. I. principles and methods, Applied Spectroscopy Reviews, 39:3, 295-363 (2004).

[8] Schwaighofer, A., Brandstetter, M. Lendl, B., Quantum cascade lasers (QCLs) in biomedical spectroscopy, Chem. Soc. Rev., 46, 5903 (2017).

[9] Akhgar, C., Ramer, G., Żbik, M., Trajnerowicz, A., Pawluczyk, J., Schwaighofer, A., Lendl, B., The next generation of IR spectroscopy: EC-QCL-based mid-IR transmission spectroscopy of proteins with balanced detection, Anal. Chem., 92, 9901−9907 (2020).

[10] https://www.gassensing.co.uk

[11] Won, R. Integrating silicon photonics. Nature Photon 4, 498–499 (2010).

[12] Lipson, M. The revolution of silicon photonics. Nat. Mater. 21, 974–975 (2022).

[13] Hu, T., Dong, B., Luo, X., Liow, T-Y., Song, J., Lee, C. Lo, G-Q., Silicon photonic platforms for mid-infrared applications, Photon. Res. 5, 417-430 (2017).

[14] Soref, R., Mid-infrared photonics in silicon and germanium, Nature Photonics, 4, (2010).

[15] Izawa, T. , Shibata, N., Takeda, A., Optical attenuation in pure and doped fused silica in the ir wavelength region, Appl. Phys. Lett. 31, 33 (1977).

[16] Efimov, A.M., Pogareva, V.G., IR absorption spectra of vitreous silica and silicate glasses: The nature of bands in the 1300 to 5000 cm−1 region, Chemical Geology, 229, 1–3, 198 (2006).

[17] Mashanovich, G. Z. , Nedeljkovic, M., Soler-Penades, J., Qu, Z., Cao, W., Osman, A., Wu, Y., Stirling, C. J., Qi, Y., Cheng, Y.X., Reid, L., Littlejohns, C. G. , Kang, J. , Zhao, Z. , Takenaka, M. , Li, T. , Zhou, Z., Gardes, F. Y., Thomson, D. J., Reed, G. T. , Group IV mid-infrared photonics, Opt. Mater. Express 8, 2276-2286 (2018).

[18] I. A. Fischer, M. Brehm, M. De Seta, G. Isella, D. J. Paul, M. Virgilio, G. Capellini; On-chip infrared photonics with Si-Ge-heterostructures: What is next?. *APL Photonics*; 7 (5): 050901 (2022).

[19] Marris-Morini, D., Vladyslav, V., Ramirez, J-M., Liu, Q., Ballabio, A., Frigerio, J., Montesinos, M., Alonso-Ramos, C., Le Roux, X., Serna, S., Benedikovic, D., Chrastina, D., Vivien, L., Isella, G., Germanium-based integrated photonics from near- to mid-infrared applications, Nanophotonics, 7, 11, 1781-1793 (2018).

[20] Reboud, V., Concepción, O., Du, W. , El Kurdi, M., Hartmann, J.M., Ikonic, Z. , Assali, S., Pauc, N., Calvo, V. , Cardoux, C., Kroemer, E. , Coudurier, N., Rodriguez, P. , Yu, S.-Q., Buca, D., Chelnokov, A., Advances in GeSn alloys for MIR applications, Photonics and Nanostructures - Fundamentals and Applications 58, 101233 (2024).

[21] Johnson, F.A., Lattice absorption bands in silicon, Proc. Phys. Soc. 73 265 (1959).

[22] Fray, S.J., Johnson, F.A., Quarrington, J.E., Williams, N., Lattice bands in germanium, *Proc. Phys. Soc.* 85 153 (1965).

[23] Hawkins, G., Hunneman, R., The temperature-dependent spectral properties of filter substrate materials in the far-infrared (6–40 µm), Infrared Physics & Technology, 45(1),69-79 (2004)

[24] Lee, Y.-J., Das, A., Mah, M. L., Talghader, J. J. Long-wave infrared absorption measurement of undoped germanium using photothermal common-path interferometry. Appl. Opt. 59, 3494 (2020).

[25] Tetzner, H., Corley-Wicia, A. A., Devaiya, A. J., Concepción, O., Stolarek, D., Schubert, M. A., Yamamoto, Y. , Buca, D., Capellini, G., Dislocations influence the background hole densities in Ge/Si virtual substrates, Appl. Phys. Lett. 127, 251901 (2025).

[26] Nedeljkovic, M., Soref, R., Mashanovich, G.Z., Predictions of free-carrier electroabsorption and electrorefraction in germanium, IEEE Photonics Journal, vol. 7, 2419217 (2015).

[27] Pesarcik, S. F., Treyz, G. V., Iyer, S. S. & Halbout, J. M., Silicon germanium optical waveguides with 0-5 dB/cm losses for singlemode fibre optic systems. Electron. Lett. 28, 159 (1992).

[28] Soref, R.A, Emelett, S.J., Buchwald, W.R., Silicon waveguided components for the long-wave infrared region, J. Opt. A: Pure Appl. Opt. 8 840 (2006)

[29] Michel, J., Liu, J. & Kimerling, L. C. High-performance Ge-on-Si photodetectors. Nature Photonics 4, 527–534 (2010).

[30] Nedeljkovic, M., Soler Penades, J., Mittal,V., Senthil Murugan,G., Khokhar, A.Z., Littlejohns, C., Carpenter, L.G., Gawith, C.B.E., Wilkinson, J.S., Mashanovich, G.Z., Germanium-on-silicon waveguides operating at mid-infrared wavelengths up to 8.5 µm. Opt. Express 25, 27431 (2017).

[31] Mittal, V., Devitt, G., Nedeljkovic, M., Carpenter, L. G., Chong, H. M. H., Wilkinson, J. S. , Mahajan, S., Mashanovich, G.Z., Ge on Si waveguide mid-infrared absorption spectroscopy of proteins and their aggregates, Biomedical Optics Express, 11, 4714-4722, (2020).

[32] Shen, L., Healy, N., Mitchell, C.J., Soler Penades, J., Nedeljkovic, M., Mashanovich, G.Z., Peacock, A.C., Mid-infrared all-optical modulation in low-loss germanium-on-silicon waveguides, Opt. Lett. 40, 268-271 (2015).

[33] Nedeljkovic, M., Penades, J. S., Mitchell, C. J., Khokhar, A. Z., Stankovic, S., Bucio, T. D., Littlejohns, C. G., Gardes, F. Y., Mashanovich, G. Z., Surface-grating-coupled low-loss Ge-on-Si rib waveguides and multimode interferometers," IEEE Photon. Technol. Lett. 27, 1040–1043 (2015)

[34] Della Torre, A., Sinobad, M., Armand, R., Luther-Davies, B., Ma, P., Madden, S., Mitchell, A., Moss, D. J., Hartmann, J.-M., Reboud, V., Fedeli, J.-M., Monat, C., Grillet, C., Mid-infrared supercontinuum generation in a low-loss germanium-on-silicon waveguide, APL Photon. 6, 016102 (2021).

[35] Morgan, R., Heidelberger, C., Kharas, D., Mahony, T., Sumant, A., Cahoy, K., Sorace-Agaskar, C., Development and characterization of low-loss waveguides and top claddings for germanium-on-silicon mid-wave infrared integrated photonics, Opt. Express 33, 44309 (2025).

[36] Kozak, D.A., Tyndall, N.F., Pruessner, M.W., Rabinovich, W.S., Stievater, T.H., Germanium-on-silicon waveguides for long-wave integrated photonics: ring resonance and thermo-optics, Opt. Express 29, 15443 (2021).

[37] Gallacher, K., Millar, R.W., Griškevičiūte, U., Baldassarre, L., Sorel, M., Ortolani, M., Paul, D. J., Low loss Ge-on-Si waveguides operating in the 8–14 µm atmospheric transmission window, Opt. Express 26, 25667-25675 (2018).

[38] Littlejohns, C.G.; Rowe, D.J.; Du, H.; Li, K.; Zhang, W.; Cao, W.; Dominguez Bucio, T.; Yan, X.; Banakar, M.; Tran, D.; et al. CORNERSTONE's silicon photonics rapid prototyping platforms: current status and future outlook. *Appl. Sci. 10*, 8201 (2020).

[39] Teigell Benéitez, N., Baumgartner, B., Missinne, J., Radosavljevic, S., Wacht, D., Hugger, S., Leszcz, P., Lendl, B., Roelkens, G., Mid-IR sensing platform for trace analysis in aqueous solutions based on a germanium-on-silicon waveguide chip with a mesoporous silica coating for analyte enrichment, Optics Express, 28, 18, 27013, (2020).

[40] Radosavljevic, S., Teigell Beneitez, N., Katumba, A., Muneeb, M., Vanslembrouck, M., Kuyken, B., Roelkens, G., Mid-infrared Vernier racetrack resonator tunable filter implemented on a germanium on SOI waveguide platform, Opt. Mater. Express 8, 824-835 (2018)

[41] Chandler-Horowitz, D., Amirtharaj, P.M., High-accuracy, midinfrared refractive index values of silicon. J. Appl. Phys., 97 (12), 123526 (2005).

[42] Shkondin, E., Takayama, O., Aryaee Panah, M. E., Liu, P. , Larsen, P. V., Mar, M. D., Jensen, F., Lavrinenko, A.V., Large-scale high aspect ratio Al-doped ZnO nanopillars arrays as anisotropic metamaterials, Opt. Mater. Express 7, 1606-1627 (2017).

[43] Sinobad, M., Monat, C., Luther-davies, B., Ma, P., Madden, S., Moss, D.J., Mitchell, A., Allioux, D., Orobtchouk, R., Boutami, S., Hartmann J-M., Fedeli, J-M., Grillet, C., Mid-infrared octave spanning supercontinuum generation to 8.5 µm in silicon-germanium waveguides," Optica 5, 360-366 (2018).

[44] Brun, M. , Labeye, P., Grand, G., Hartmann, J-M., Boulila, F., Carras, M., Nicoletti, S., Low loss SiGe graded index waveguides for mid-IR applications. Opt. Express 22, 508 (2014).

[45] Currie, M. T., Samavedam, S. B., Langdo, T. A., Leitz, C. W. & Fitzgerald, E. A. Controlling threading dislocation densities in Ge on Si using graded SiGe layers and chemical-mechanical polishing. Appl. Phys. Lett. 72, 1718–1720 (1998).

[46] Montesinos Ballester, M., Vakarin, V., Liu, Q., Le Roux, X., Frigerio, J., Ballabio, A., Barzaghi, A., Alonso-Ramos, C., Vivien, L., Isella, G., Marris-Morini, D., Ge-rich graded SiGe waveguides and interferometers from 5 to 11 µm wavelength range," Opt. Express 28, 12771-12779 (2020)

[47] Turpaud V., Nguyen, T-H-N., Dely, H., Koompai, N., Bricout, A., Hartmann, J-M., Bernier, N., Krawczyk, J., Lima, G., Edmond, S., Herth, E., Alonso-Ramos, C., Vivien, L., Marris-Morini, D., Low-loss SiGe waveguides for mid-infrared photonics fabricated on 200 mm wafers, Optics Express **32**, 17400-17408 (2024)

[48] Deguet, D., Morales, C., Dechamp, J., Hartmann, J. M., Charvet, A. M., Moriceau, H., Chieux, F., Beaumont, A., Clavelier, L., Loup, V., Kernevez, N., Raskin, G., Richtarch, C., Allibert, F., Letertre, F., Mazure, C., Germanium-on-insulator (GeOI) structures realized by the Smart Cut™ technology,*2004 IEEE International SOI Conference (IEEE Cat. No.04CH37573)*, Charleston, SC, USA, 96-97, doi: 10.1109/SOI.2004.1391571,(2004).

[49] Zhao, Z., Lim, C.-M., Ho, C., Sumita, K. , Miyatake, Y. , Toprasertpong, K., Takagi, S., Takenaka, M., Low-loss Ge waveguide at the 2-µm band on an n-type Ge-on-insulator wafer, Optical Materials Express Vol. 11, Issue 12, pp. 4097-4106 (2021).

[50] Osman, A., Nedeljkovic, M., Soler Penades, J., Wu, Y., Qu, Z., Khokhar, A. Z., Debnath, K. , Mashanovich, G. Z. , Suspended low-loss germanium waveguides for the longwave-infrared, Optics Letters, vol. 43, pp. 5997-6000, 2018.

[51] Kim, SH. , Han, J-H., Shim, J-P., Kim, H-J., Choi, W-J., Verification of Ge-on-insulator structure for a mid-infrared photonics platform," Opt. Mater. Express 8, 440-451 (2018)

[52] Ren, D., Dong, C., Høvik, J., Khan, M., Aksnes, A., Fimland, B. , Burghoff, D. "Low-loss hybrid germanium-on-zinc selenide waveguides in the longwave infrared" *Nanophotonics*, vol. 13, no. 10, 2024, pp. 1815-1822.

[53] Sánchez-Postigo, A., Ortega-Moñux, A., Soler Penadés, J., Osman, A., Nedeljkovic, M., Qu, Z., Wu, Y., Molina-Fernández, I., Cheben, P., Mashanovich, G.Z., Wangüemert-Pérez, J.G., Suspended germanium waveguides with subwavelength-grating metamaterial cladding for the mid-infrared band, Opt. Express 29, 16867 (2021)

[54] Soler Penades, J., Ortega-Moñux, A. , Nedeljkovic, M., Wangüemert-Pérez, J. G., Halir, R., Khokhar, A.Z., Alonso-Ramos, C., Qu, Z., Molina-Fernández, I., Cheben, P., Mashanovich, G.Z., Suspended silicon mid-infrared waveguide devices with subwavelength grating metamaterial cladding, Optics Express, vol. 24, 22908 (2016).

[55] Rytov, S., Electromagnetic properties of a finely stratified medium, Soviet Physics JEPT 2, 466–475 (1956)

[56] Mait J. N., Prather, D. W. Selected Papers on Subwavelength Diffractive Optics, SPIE Book (2001).

[57] Lalanne, P., Astilean, S., Chavel, P., Cambril, E., Launois, H., Design and fabrication of blazed binary diffractive elements with sampling periods smaller than the structural cutoff, J. Opt. Soc. Am. A 16, 1143-1156 (1999)

[58] Kuznetsov, A. I., Miroshnichenko, A. E., Brongersma, M. L. , Kivshar, Y.S., Luk'Yanchuk B., Optically resonant dielectric nanostructures, Science 354, 6314 (2016).

[59] Halir, R., Bock, P.J., Cheben, P., Ortega-Moñux, A., Alonso-Ramos, C., Schmid, J.H., Lapointe, J., Xu, D.-X., Wangüemert-Pérez, J.G., Molina-Fernández, Í. and Janz, S., Waveguide sub-wavelength structures: a review of principles and applications. Laser & Photonics Reviews, 9: 25-49 (2015).

[60] Cheben, P., Halir, R., Schmid, J.H. et al. Subwavelength integrated photonics. Nature 560, 565–572 (2018)

[61] Cheben, P., Schmid, J.H., Halir, R., Luque-González, J.M., Wangüemert-Pérez, J.G., Melati, D., Alonso-Ramos, C., Recent advances in metamaterial integrated photonics, Adv. Opt. Photon. 15, 1033-1105 (2023).

[62] Cheben, P. , Xu, D-X., Janz, S., Densmore, A., Subwavelength waveguide grating for mode conversion and light coupling in integrated optics," Opt. Express 14, 4695–4702 (2006).

[63] Schmid, J.H., Cheben, P., Janz, S., Lapointe, J., Post, E., Xu, D.-X., Gradient-index antireflective subwavelength structures for planar waveguide facets, Opt. Lett. 32, 1794-1796 (2007)

[64] Levy, U., Abashin,M., Ikeda, K., Krishnamoorthy, A., Cunningham, J., Fainman, Y., Inhomogenous dielectric metamaterials with space-variant polarizability, Phys. Rev. Lett. 98, 243901 (2007).

[65] Alonso-Ramos, C., Nedeljkovic, M., Benedikovic, D., Soler Penadés, J., Littlejohns, C., Pérez-Galacho, D., Vivien, L., Cheben, P., Mashanovich, G. Z., Germanium-on-silicon mid-infrared grating couplers with low-reflectivity inverse taper excitation, Optics Letters, 41, 4324-4327 (2016).

[66] Sánchez-Postigo, A., Ortega-Moñux, A., Pereira-Martín, D., Molina-Fernández, Í., Halir, R. , Cheben, P. , Soler Penadés, J. , Nedeljkovic, M. , Mashanovich, G. Z. , Wangüemert-Pérez, J.G., Design of a suspended germanium micro-antenna for efficient fiber-chip coupling in the long-wavelength mid-infrared range, Optics Express, 27, 22302 (2019).

[67] Perestjuk, M., Armand, R., Sandoval Campos, M., Ferhat, L., Reboud, V., Bresson, N., Hartmann, J-M., Mathieu, V., Ren, G., Boes, A., Mitchell, A., Monat, C., Grillet, C. One million quality factor integrated ring resonators in the mid-infrared, Nanophotonics, vol. 14, no. 7, 1009-1015 (2025).

[68] Ren, D., Dong, C., Addamane, S.J. et al. High-quality microresonators in the longwave infrared based on native germanium. Nat Commun 13, 5727 (2022).

[69] Koompai, N., Nguyen, T. H. N., Turpaud, V., Frigerio, J., Falcone, V., Calcaterra, S., Lucia, L., Bousseksou, A., Colombelli, R., Coudevylle, J.-R., Bouville, D., Alonso-Ramos, C., Vivien, L., Isella, G.,

Marris-Morini, D., Long-wave infrared integrated resonators in the 7.5–9 µm wavelength range, Appl. Phys. Lett. 123, 031109 (2023)

[70] Xiao, T.-H., Zhao, Z., Zhou, W. , Takenaka, M., Ki Tsang, H., Cheng, Z., Goda, K. , High-Q germanium optical nanocavity, Photonics Research, 6, 925 (2018).

[71] Liu, Q., Ramirez, J-M., Vakarin, V., Le Roux, X. , Frigerio, J., Ballabio, A., Talamas Simola, E., Alonso-Ramos, C., Benedikovic, D., Bouville, D. , Vivien, L., Isella, G., Marris-Morini, D. , On-chip Bragg grating waveguides and Fabry- Perot resonators for long-wave infrared operation up to 8.4 µm, Optics Express, 26, 26, 34366 (2018).

[72] Malik, A., Muneeb, M. , Pathak, S. , Shimura, Y., Van Campenhout, J., Loo, R. , Roelkens, G., Germanium-on-Silicon Mid-Infrared Arrayed Waveguide Grating Multiplexers, IEEE Photonics Technology Letters, 25, 18, 1805 (2013).

[73] Malik, A., Muneeb, M., Shimura, Y., Van Campenhout, J., Loo, R., Roelkens, G., Germanium-on-silicon planar concave grating wavelength (de)multiplexers in the mid-infrared, Applied Physics Letters, 103, 161119 (2013).

[74] Montesinos-Ballester, M., Liu, Q., Vakarin, V., Manel Ramirez, J., Alonso-Ramos, C., Le Roux, X., Frigerio, J., Ballabio, A., Talamas, E., Vivien, L., Isella, G., Marris-Morini, D., On-chip Fourier-transform spectrometer based on spatial heterodyning tuned by thermo-optic effect, Scientific Reports 9, 14633 (2019).

[75] https://www.agilent.com/en/product/molecular-spectroscopy/ftir-spectroscopy/ftir-compact-portable-systems/5500-series-compact-ftir

[76] Brès, C-S., Della Torre, A., Grassani, D., Brasch, V., Grillet, C., Monat, C., Supercontinuum in integrated photonics: generation, applications, challenges, and perspectives, Nanophotonics, 12, 7, 1199 (2023).

[77] Hon, N.K., Soref, R., Jalali, B., The third-order nonlinear optical coefficients of Si, Ge, and $Si_{1-x}Ge_x$ in the midwave and longwave infrared, J. Appl. Phys. , 110, 1, 011301 (2011).

[78] Serna, S., Vakarin, V., Ramirez, JM. et al. Nonlinear Properties of Ge-rich $Si_{1-x}Ge_x$ materials with different Ge concentrations. Sci Rep 7, 14692 (2017).

[79] Ettabib, M.A., Xu, L., Bogris, A., Kapsalis, A., Belal, M., Lorent, E., Labeye, P., Nicoletti, S., Hammani, K., Syvridis, D., Shepherd, D.P., Price, J.H.V., Richardson, D.J., Petropoulos, P., Broadband telecom to mid-infrared supercontinuum generation in a dispersion-engineered silicon germanium waveguide, Opt. Lett. 40, 4118-4121 (2015).

[80] Sinobad, M., DellaTorre, A., Armand, R., Luther-Davies, B., Ma, P., Madden, S., Mitchell, A., Moss, D.J., Hartmann, J-M., Fedeli, J-M., Monat, C., Grillet, C., Mid-infrared supercontinuum generation in silicon-germanium all-normal dispersion waveguides, Opt. Lett. 45, 5008-5011 (2020).

[81] Sinobad, M., Della Torre, A., Armand, R., Luther-Davies, B., Ma, P., Madden, S., Mitchell, A., Moss, D.J., Hartmann, J-M., Fédéli, J-M., Monat, C., Grillet, C., High coherence at f and 2f of mid-infrared supercontinuum generation in silicon germanium waveguides, *IEEE Journal of Selected Topics in Quantum Electronics*, 26, 2, 1 (2020).

[82] Sinobad, M., Della Torre, A., Luther-Davis, B., Ma, P., Madden, S., Debbarma, S., Vu, K., Moss, D.J., Mitchell, A., Hartmann, J-M., Fedeli, J-M., Monat, C., Grillet, C., Dispersion trimming for mid-infrared

supercontinuum generation in a hybrid chalcogenide/silicon-germanium waveguide, J. Opt. Soc. Am. B 36, A98-A104 (2019).

[83] Montesinos-Ballester, M., Lafforgue, C., Frigerio, J., Ballabio, A., Vakarin, V., Liu, Q., Ramirez, J-M., Le Roux, X., Bouville, D., Barzaghi, A., Alonso-Ramos, C., Vivien, L., Isella, G., Marris-Morini, D., On-chip mid-infrared supercontinuum generation from 3 to 13 µm wavelength, ACS Photonics 7 (12), 3423 (2020).

[84] Della Torre, A. Armand, R., Sinobad, M., Fiaboe K.F., Luther-Davies B., Madden S.J., Mitchell A., Nguyen T., Moss D.J., Hartmann, J-M., Reboud, V., Fédéli, J-M., Monat, C., Grillet C., Mid-Infrared supercontinuum generation in a varying dispersion waveguide for multi-species gas spectroscopy," in *IEEE Journal of Selected Topics in Quantum Electronics*, vol. 29, no. 1: Nonlinear Integrated Photonics (2023).

[85] Wang, F.; Qi, X.; Chen, Z.; Razeghi, M.; Dhillon, S. Ultrafast pulse generation from quantum cascade lasers. Micromachines 13, 2063 (2022).

[86] Hillbrand, J., Opačak, N., Piccardo, M. et al. Mode-locked short pulses from an 8 µm wavelength semiconductor laser. Nat Commun 11, 5788 (2020)

[87] Täschler, P., Bertrand, M., Schneider, B. et al. Femtosecond pulses from a mid-infrared quantum cascade laser. Nat. Photon. 15, 919–924 (2021).

[88] Täschler, P., Miller, L., Kapsalidis, F., Beck, M., Faist, J., Short pulses from a gain-switched quantum cascade laser, Optica 10, 507-512 (2023).

[89] Bricout, A., Bertrand, A., Täschler, P., Schneider, B., Turpaud, V., Calcaterra, S., Impelluso, D., Faverzani, M., Bouville, D., Coudevylle, J-R., Edmond, S., Herth, E., Alonso-Ramos, C., Vivien, L., Frigerio, J., Isella, G., Faist, J., Marris-Morini, D., Pulse Generation by On-Chip Dispersion Compensation at 8 µm Wavelength Laser & Photonics Reviews, 0:e01485 (2025)

[90] Bagheri, M., Frez, C., Sterczewski, L.A. et al. Passively mode-locked interband cascade optical frequency combs. Sci Rep 8, 3322 (2018)

[91] Hillbrand, J., Beiser, M., Andrews, A.M., Detz, H., Weih, R., Schade, A., Höfling, S., Strasser, G., Schwarz, B., Picosecond pulses from a mid-infrared interband cascade laser, Optica 6, 1334-1337 (2019)

[92] Turpaud, V., Koompai, N., Nguyen, T.H.N., Yang, Y., Frigerio, F., Coudevylle, J.R., Bouville, D., Alonso-Ramos, C., Herth, E., Vivien, L., Isella, G., Marris-Morini, D., Enhancing Mid-Infrared Supercontinuum generation at low pump power in SiGe waveguides, *2023 Conference on Lasers and Electro-Optics Europe & European Quantum Electronics Conference (CLEO/Europe-EQEC)*, Munich, Germany, 2023, doi: 10.1109/CLEO/Europe-EQEC57999.2023.10232668.

[93] Elbaz, A., Buca, D., von den Driesch, N. Pantzas, K., Patriarche, G, Zerounian, N., Herth, E., Checoury, X., Sauvage, S., Sagnes, I., Foti, A. , Ossikovski, R., Hartmann, J-M., Boeuf, F., Ikonic, Z., Boucaud, P., Grützmacher, D., El Kurdi, M.,. Ultra-low-threshold continuous-wave and pulsed lasing in tensile-strained GeSn alloys. Nat. Photonics 14, 375–382 (2020).

[94] Zhou, Y., Dou, W., Du, W., Ojo, S., Tran, H., Ghetmiri, S.A., Liu, J., Sun, G., Soref, R., Margetis, J., Tolle, J., Li, B., Chen, Z., Mortazavi, M., Yu, S-Q., Optically pumped GeSn lasers operating at 270 K with broad waveguide structures on Si, ACS Photonics 6 (6), 1434 (2019).

[95] Zhou, Y., Ojo, S., Wu, C-W., Miao, Y., Tran, H., Grant, J.M., Abernathy, G., Amoah, S., Bass, J., Salamo, G., Du, W., Chang, G-E., Liu, J., Margetis, J., Tolle, J., Zhang, Y-H., Sun, G., Soref, R.A., Li, B., Yu, S-Q., Electrically injected GeSn lasers with peak wavelength up to 2.7 µm, Photon. Res. 10, 222-229 (2022)

[96] Marzban, B., Seidel, L., Liu, T., Wu, K., Kiyek, V., Hartwig Zoellner, M., Ikonic, Z., Schulze, J., Grützmacher,D., Capellini, G., Oehme, M., Witzens, J., Buca, D., Strain Engineered Electrically Pumped SiGeSn Microring Lasers on Si, ACS Photonics 10 (1), 217-224 (2023).

[97] Tournié, E., Monge Bartolome, L., Rio Calvo, M. , Loghmari, Z., Díaz-Thomas, D. A., Teissier, R., Baranov, A. N., Cerutti, L., Rodriguez, J-B., Mid-infrared III–V semiconductor lasers epitaxially grown on Si substrates. Light Sci Appl 11, 165 (2022).

[98] Mitchell, C.J., Osman, A., Li, K., Panadès, J.S., Nedeljkovic, M., Zhou, L., Groom, K.M., Heffernan, J., Mashanovich, G., Hybrid laser integration in the mid-IR for silicon photonics sensing applications, Proc. SPIE 12426, Silicon Photonics XVIII, 1242608 (13 March 2023); doi: 10.1117/12.265005

[99] Park, J. S. Lin, T.L., Jones, E.W., Del Castillo, H.M., Gunapala, S.D., Long-wavelength stacked SiGe/Si heterojunction internal photoemission infrared detectors using multiple SiGe/Si layers, Appl. Phys. Lett. 64, 2370 (1994).

[100] Boucaud P., Gauthier-Lafaye, O., Lourtioz, J-M., Julien F.H., Dekel E., Ehrenfreund E., Gershoni D., Sagnes I., Campidelli, Y., Absorption and emission spectroscopy of intersubband transitions in SiGe/Si quantum wells , J. Vac. Sci. Technol. B 16, 1697 (1998).

[101] Nguyen, T.H.N., Koompai, N., Turpaud, V., Montesinos-Ballester, M., Frigerio, J., Calcaterra, S., Ballabio, A., Le Roux, X., Coudevylle, J.R., Villebasse, C., Bouville, D., Alonso-Ramos, C., Vivien, L., Isella, G. and Marris-Morini, D. Room temperature-integrated photodetector between 5 µm and 8 µm wavelength. Adv. Photonics Res., 4: 2200237 (2023).

[102] Nguyen, T. H. N., Turpaud, V., Koompai, N., Peltier, J., Calcaterra, S., Isella, G., Coudevylle, J-R., Alonso-Ramos, C., Vivien, L., Frigerio, J. , Marris-Morini, D., Integrated PIN modulator and photodetector operating in the mid-infrared range from 5.5 µm to 10 µm, Nanophotonics, (2024)

[103] Stewart, J.W., Wilson, N.C., Mikkelsen, M.H., Nanophotonic Engineering: A new paradigm for spectrally sensitive thermal photodetectors, ACS Photonics 8 (1), 71-84 (2021).

[104] Shim, J., Lim, J., Kim, I. et al. Room-temperature waveguide-integrated photodetector using bolometric effect for mid-infrared spectroscopy applications. Light Sci Appl 14, 125 (2025).

[105] Huang, G.Y., Hao, Y., Li, S.Q., Jia, Y.D., Guo,J.C., Zhang, H., Wang, B., Recent progress in waveguide-integrated photodetectors based on 2D materials for infrared detection, Journal of Physics D: Applied Physics, Volume 56, Number 11 (2023)

[106] Li,K., Liu, S., Thomson, D.J., Zhang, W., Yan, X., Meng, F., Littlejohns, C.G., Du, H., Banakar, M., Ebert, M., Cao, W., Tran, D., Chen, B., Shakoor, A., Petropoulos, P., Reed, G.T., Electronic–photonic convergence for silicon photonics transmitters beyond 100 Gbps on–off keying, Optica 7, 1514-1516 (2020).

[107] Xu, T., Dong, Y., Zhong, Q., Zheng, S., Qiu, Yang, Z., Xingyan, J., Lianxi, L., Cheng K. and Hu, T., Mid-infrared integrated electro-optic modulators: a review, Nanophotonics, 12, 19, 3683 (2023).

[108] Malik, A., Dwivedi, S., Van Landschoot, L., Muneeb, M., Shimura, Y., Lepage, G., Van Campenhout, J., Vanherle, W., Van Opstal, T., Loo, R., Roelkens, D., Ge-on-Si and Ge-on-SOI thermo-optic phase shifters for the mid-infrared, Opt. Express 22, 28479-28488 (2014).

[109] Nedeljkovic, M., Soref R., Mashanovich, G.Z., Free-carrier electrorefraction and electroabsorption modulation predictions for silicon over the 1–14- μm infrared wavelength range, in IEEE Photonics Journal, 3, 6, 1171 (2011).

[110] Li, T., Nedeljkovic, M., Hattasan, N., Cao, W., Qu, Z., Littlejohns, C.G., Soler Penades, J., Mastronardi, L., Mittal, V., Benedikovic, D., Thomson, D.J., Gardes, F.Y., Wu, H., Zhou, Z., Mashanovich, G.Z., Ge-on-Si modulators operating at mid-infrared wavelengths up to 8 μm, Photon. Res. 7, 828-836 (2019).

[111] Nguyen, T.H.N., Koompai, N., Turpaud, V., Montesinos-Ballester, M., Peltier, J., Frigerio, J., Ballabio, A., Giani, R., Coudevylle, J-R., Villebasse, C, Bouville, D., Alonso-Ramos, C., Vivien, L., Isella, G., Marris-Morini, D., 1 GHz electro-optical silicon-germanium modulator in the 5-9 µm wavelength range, Opt. Express 30, 47093-47102 (2022).

[112] Barzaghi, A., Falcone, V., Calcaterra, S., Marris-Morini, D., Virgilio, M., Frigerio, J., Modelling of an intersubband quantum confined Stark effect in Ge quantum wells for mid-infrared photonics, Opt. Express 30, 46710-46721 (2022)

[113] Akie, M., Fujisawa, T., Sato, T., Arai, M., Saitoh, K., GeSn/SiGeSn multiple-quantum-well electroabsorption modulator with taper coupler for mid-infrared Ge-on-Si platform, IEEE Journal of Selected Topics in Quantum Electronics, 24, 6, 1 (2018).

[114] Yamaguchi, Y., Takagi, S., Takenaka, M., Low-loss graphene-based optical phase modulator operating at mid-infrared wavelength, Jpn. J. Appl. Phys. 57 (2018).

[115] Li, Q., Xiong, Z., Yan, Z., Cheng, G., Xu, F., Shen, Z., Yi, Q., Yu, Y., Shen, L, High-speed mid-infrared graphene electro-optical modulator based on suspended germanium slot waveguides, Opt. Express 31, 29523-29535 (2023)

[116] Fortier, T., Baumann, E. 20 years of developments in optical frequency comb technology and applications. *Commun Phys* **2**, 153 (2019).

[117] Levy, J., Gondarenko, A., Foster, M. et al. CMOS-compatible multiple-wavelength oscillator for on-chip optical interconnects. Nature Photon 4, 37–40 (2010).

[118] Gaeta, A.L., Lipson, M. , Kippenberg, T.J. Photonic-chip-based frequency combs. Nature Photon 13, 158–169 (2019).

[119] Helgason, Ó.B., Girardi, M., Ye, Z. Lei, F., Schröder, J., Torres-Company, V., Surpassing the nonlinear conversion efficiency of soliton microcombs, Nat. Photon. 17, 992–999 (2023).

[120] Yu, M., Okawachi, Y., Griffith, A.G., Lipson, M., Gaeta, A.L., Mode-locked mid-infrared frequency combs in a silicon microresonator," Optica 3, 854-860 (2016)

[121] Luke, K, Okawachi, Y., Lamont, M.R.R., Gaeta, A.L., Lipson, M., Broadband mid-infrared frequency comb generation in a $Si_3N_4$ microresonator, Opt. Lett. 40, 4823-4826 (2015)

[122] Griffith, A., Lau, R., Cardenas, J., Okawachi, Y., Mohanty, A., Fain, R., Lee, Y.H.D, Yu, M., Phare, C.T., Poitras, C.B., Gaeta, A.L., Lipson, M., Silicon-chip mid-infrared frequency comb generation. Nat Commun 6, 6299 (2015).

[123] Parriaux, A., Hammani, K., Millot, G., Electro-optic frequency combs, Adv. Opt. Photon. 12, 223-287 (2020).

[124] Turpaud, V., Nguyen, T.-H-N., Koompai, N., Peltier, J., Frigerio, J., Calcaterra, S., Coudevylle, J-R., Bouville, D., Alonso-Ramos, C., Vivien, L., Isella, G., Marris-Morini, D., Tunable on-chip electro-optic frequency-comb generation at 8 µm wavelength, Laser and Photonics Review, 2300961 (2024).

[125] Wağli, P., Chang, Y-C., Homsy, A., Hvozdara, L., Herzig, H.P., de Rooij, N.F., Microfluidic droplet-based liquid−liquid extraction and on-chip IR spectroscopy detection of cocaine in human saliva , Anal. Chem., 85, 7558 (2013).

[126] Mittal, V., Nedeljkovic, M., Carpenter, L.G., Khokhar, A.Z., Chong, H.M.H., Mashanovich, G.Z., Bartlett, P.N., Wilkinson, J.S., Waveguide absorption spectroscopy of bovine serum albumin in the mid-infrared fingerprint region, ACS Sens., 4, 1749−1753 (2019).

[127] David, M. , Doganlar, I. C., Nazzari, D. , Arigliani, E., Wacht, D. , Sistani, M., Detz, H., Ramer, G., Lendl, B., Weber, W. M., Strasser, G. , Hinkov, B. , Surface protection and activation of mid-IR plasmonic waveguides for spectroscopy of liquids, Journal of Lightwave Technology, 42, 2, 748-759, (2024)

[128] Picqué, N., Hänsch, T.W. Frequency comb spectroscopy. Nature Photon 13, 146–157 (2019).

[129] D. Skryabin et al, Roadmap: Emerging Platforms and Applications of Optical Frequency Combs and Dissipative Solitons, https://arxiv.org/pdf/2511.18231